\documentclass{emulateapj}
\usepackage{apjfonts}
\usepackage{epsf}
\begin{document}

\slugcomment{Submitted to ApJL}
\shortauthors{J. M. Miller et al.}
\shorttitle{A NuSTAR Observation of Serpens X-1}

\title{Constraints on the Neutron Star and Inner Accretion Flow in Serpens X-1 Using NuSTAR}

\author{J.~M.~Miller\altaffilmark{1},
  M. L. Parker\altaffilmark{2}, F. Fuerst\altaffilmark{3},
  M. Bachetti\altaffilmark{4,5}, D. Barret\altaffilmark{4,5},
  B. W. Grefenstette\altaffilmark{3}, S. Tendulkar\altaffilmark{3},
  F. A. Harrison\altaffilmark{3}, 
  S. E. Boggs\altaffilmark{6},
  D. Chakrabarty\altaffilmark{7},
  F. E. Christensen\altaffilmark{8},
  W. W. Craig\altaffilmark{9,10},
  A. C. Fabian\altaffilmark{2},
  C. J. Hailey\altaffilmark{10},
  L. Natalucci\altaffilmark{11}, 
  F. Paerels\altaffilmark{10}, 
  V. Rana\altaffilmark{3}, 
  D. K. Stern\altaffilmark{12},
  J. A. Tomsick\altaffilmark{6},
  W. W. Zhang\altaffilmark{13}
}

\altaffiltext{1}{Department of Astronomy, The University of Michigan, 500
Church Street, Ann Arbor, MI 48109-1046, USA, jonmm@umich.edu}

\altaffiltext{2}{Institute of Astronomy, The University of Cambridge,
  Madingley Road, Cambridge, CB3 OHA, UK}

\altaffiltext{3}{Cahill Center for Astronomy and Astrophysics,
  California Institute of Technology, Pasadena, CA, 91125 USA}

\altaffiltext{4}{Universite de Toulouse, UPS-OMP, Toulouse, France}

\altaffiltext{5}{CNRS, Institut de Recherche en Astrophysique et
  Planetologie, 9 Av. colonel Roche, BP 44346, F-31028, Toulouse cedex
  4, France}

\altaffiltext{6}{Space Sciences Laboratory, University of California,
  Berkeley, CA 94720, USA}

\altaffiltext{7}{Kavli Institute for Astrophysics and Space Research,
  Massachusetts Institute of Technology, 70 Vassar Street, Cambridge,
  MA 02139, USA}

\altaffiltext{8}{Danish Technical University, Lyngby, DK}

\altaffiltext{9}{Lawrence Livermore National Laboratory, Livermore, CA}

\altaffiltext{10}{Columbia Astrophysics Laboratory and Department of
  Astronomy, Columbia University, 550 West 120th Street, New York, NY
  10027, USA}

\altaffiltext{11}{Istituto di Astrofisica e Planetologia Spaziali
  (INAF), Via Fosso del Cavaliere 100, Roma, I-00133, Italy}

\altaffiltext{12}{Jet Propulsion Laboratory, California Institute of
  Technology, 4800 Oak Grove Drive, Pasadena, CA 91109, USA}

\altaffiltext{13}{NASA Goddard Space Flight Center, Greenbelt, MD 20771, USA}


\begin{abstract}
We report on an observation of the neutron star low-mass X-ray binary
Serpens X-1, made with {\em NuSTAR}.  The extraordinary sensitivity
afforded by {\em NuSTAR} facilitated the detection of a clear, robust,
relativistic Fe K emission line from the inner disk.  A relativistic
profile is required over a single Gaussian line from any charge state
of Fe at the $5\sigma$ level of confidence, and any two Gaussians of
equal width at the same confidence.  The Compton back-scattering
``hump'' peaking in the 10--20~keV band is detected for the first time
in a neutron star X-ray binary.  Fits with relativistically--blurred
disk reflection models suggest that the disk likely extends close to
the innermost stable circular orbit (ISCO) or stellar surface.  The
best-fit blurred reflection models constrain the gravitational
redshift from the stellar surface to be $z_{NS} \geq 0.16$.  The data
are broadly compatible with the disk extending to the ISCO; in that
case, $z_{NS} \geq 0.22$ and $R_{NS} \leq 12.6$~km (assuming $M_{NS} =
1.4~ M_{\odot}$ and $a=0$, where $a = cJ/GM^{2}$).  If the star is as
large or larger than its ISCO, or if the effective reflecting disk
leaks across the ISCO to the surface, the redshift constraints become
measurements.  We discuss our results in the context of efforts to
measure fundamental properties of neutron stars, and models for
accretion onto compact objects.
\end{abstract}

\section{Introduction}
The equation of state of ultradense matter remains poorly known (see,
e.g., Lattimer 2011). Such connditions cannot be replicated in
terrestrial laboratories, leaving only astrophysical studies of
neutron stars to solve this fundamental problem.  The mass-radius
relation for neutron stars depends sensitively on the equation of
state of the ultradense stellar interior. Neutron star radius
measurements are a particularly powerful discriminant, although
extreme mass measurements can also help (e.g., Demorest et al. 2010).
X-ray observations hold tremendous promise for radius measurement.
Current instruments can easily measure thermal spectra from X-ray
bursters (e.g., Guver et al. 2010) and quiescent X-ray transients
(e.g., Guillot et al. 2013), although systematic uncertainties
involving source distance, atmospheric modeling, and additional
spectral components can complicate the extraction of reliable stellar
radii (Boutloukos et al. 2010; Guver et al. 2012a, 2012b; Steiner et
al. 2013).

Whenever an accretion disk is illuminated by an external source of
hard X-rays, disk reflection is expected.  The emergent reflection
spectrum is shaped by Doppler shifts and gravitational redshifts; the
extremity of these shifts can be used to measure the location of the
inner disk with respect to the compact object (for a review, see
Miller 2007; also see Miller et al.\ 2009).  The most prominent parts
of a disk reflection spectrum are the Fe K emission line and the
Compton back-scattering "hump" that peaks between 10--30~keV.  In
principle, disk reflection spectrocopy has the potential to constrain
the radius of a neutron star, since the disk must truncate
at the surface, if not at larger radii (e.g. Cackett et al.\ 2008,
2010).

Spectra obtained with {\it XMM-Newton} and {\it Suzaku} have revealed
potentially relativistic line shapes and have started to illustrate
the power of disk reflection spectra to constrain the fundamental
parameters of neutron stars (e.g. Bhattacharyya \& Strohmayer 2007,
Cackett et al.\ 2008, 2009, 2010; Di Salvo et al.\ 2009, Iaria et
a.\ 2009, D' Ai et al.\ 2009, Papitto et al.\ 2009, Miller et
al.\ 2011, Egron et al.\ 2013).  A difficulty with X-ray CCD detectors
is that pile-up can distort the shape of Fe K lines.  Indeed, Ng et
al.\ (2010) suggested that Fe K lines in neutron stars may actually be
narrow and symmetric, and merely appear to be relativistic owing to
pile-up distortions.

However, this possibility is not borne out by simulations
(see, e.g., Miller et al.\ 2010), which instead suggest that pile-up
acts to falsely narrow lines.  Data obtained with gas spectrometers
(e.g. Lin et al.\ 2010, Cackett et al.\ 2013), which do not suffer
from photon pile-up, also point to broad lines, but such studies
are clouded by low spectral resolution.  Thus, uncertainties remain.
The detection of a Compton back-scattering ``hump'' would also help to
resolve the origin of Fe K lines in neutron stars, but no detection
has previously been achieved.

Motivated by the ability of {\it NuSTAR} (Harrison et al.\ 2013) to
reveal Fe K lines and broad-band disk reflection spectra in black
holes (e.g. NGC 1365, GRS 1915$+$105, and Cygnus X-1; Risaliti et
al.\ 2013, Miller et al.\ 2013, Tomsick et al.\ 2013), free of photon
pile-up distortions, we made a pilot observation of the well-known
neutron star X-ray binary and "atoll" source Serpens X-1.

\section{Observations and Data Reduction}
NuSTAR observed Serpens X-1 twice on 12 and 13 July, 2013 (Sequence
IDs 30001013002 and 30001013004).  The data were screened and
processed using the {\it NuSTAR} Data Analysis Software (NuSTARDAS)
version 1.1.1, resulting in a total of ~43.4 ks of on-target time.
The source is relatively bright, so the net exposure after
instrumental deadtime is taken into account is ~30.4 ks.  Spectra from
the FPMA and FPMB detectors were extracted from 120'' regions centered
on the source position, using the ``nuproducts'' FTOOL.  Corresponding
response files appropriate for the pointing type (on-axis), source
type (point), and extraction regions were also created.  Backgrounds
were created using the ``nuskybgd'' tool, which self-consistently
accounts for background variations across the face of the detector.

The spectra were analyzed using XSPEC version 12.8 (Arnaud \& Dorman
2000).  The $\chi^{2}$ statistic was used to measure the relative
quality of different spectral models.  Following Miller et
al.\ (2013), ``Churazov'' weighting was employed in all fits to govern
the influence of bins with significant but progressively less signal
at high energy (Churazov et al.\ 1996).  Errors are reported as the
$1\sigma$ confidence level on all measurements.  All fits were made
assuming a neutral equivalent neutral hydrogen column density of
$4.0\times 10^{21}~ {\rm cm}^{-2}$ (based on Dickey \& Lockman 1990),
via the ``tbabs'' model, setting ``vern'' cross sections (Verner et
al.\ 1996) and ``wilm'' abundances (Wilms et al.\ 2000).

\section{Analysis and Results}
Serpens X-1 is known to exhibit Type-1 X-ray bursts at a rate of
approximately 0.1 per hour (Galloway et al.\ 2008).  Light curves were
therefore created via the ``nuproducts'' tool and were searched for
X-ray bursts; no bursts were detected.  We also constructed an X-ray
color--color diagram using energy selections typical of diagrams made
using RXTE data, in order to provide the best correspondence.  The
{\em NuSTAR} observation was found to sample the usual "banana" branch rather
than the "island" state or any other period of extreme or unusual behavior
(see van der Klis 2006 for a description of atoll sources).

An initial inspection and comparison of the FPMA and FPMB spectra
suggested that the source is confidently detected out to 40~keV.  We
therefore restricted subsequent spectral analysis to the 3--40~keV
band.  Initial fits revealed a striking agreement between the FPMA and
FPMB spectra: when a constant was floated between the spectra to
account for uncetainties in the flux calibration of the detectors, a
value of 1.001 was measured.  We therefore created a single combined
total spectrum, background spectrum, and ancillary response matrix
using the FTOOL ``addascaspec''.  A single redistribution matrix file
was created via ``addrmf'', weighting the single responses by their
relative exposure times.  The analysis described below pertains to the
single combined spectrum.

A model consisting of a disk blackbody component (``diskbb'', see
Mitsuda et al.\ 1984), a simple single-temperature blackbody
(``bbody''), and a power-law provides a good description of
the time-averaged spectrum.  This model can be interpreted fairly
simply in terms of emission from the accretion disk, the stellar
surface, and non-thermal emission that may arise through
Comptonization and/or magnetic processes.  The simple blackbody may
not represent direct emission from the stellar surface, nor from the
entire surface.  It may represent emission that has been modified by a
boundary layer that can also be modeled via low--temperature,
optically--thick Comptonization.  However, that prescription is
similar to a hot blackbody.  Although this three-component model
describes the continuum shape fairly well, it is not a formally
acceptable fit: $\chi^{2}/\nu = 3282.8/919 = 3.57$ (also see Figures 1
and 2).

The fit is greatly improved when a relativistic line function is
added: $\chi^{2}/\nu = 1010.9/912 = 1.11$ (see Figure 1).  This model
fits the line with the ``kerrdisk'' function (Brenneman \& Reynolds
2006).  This line model was employed because it allows for small
changes to the spacetime induced by spin.  The ISCO radius changes
slowly for low values $a$, and neutron stars in X-ray binaries have
low spin parameters ($a \leq 0.3$, where $a = cJ/GM^{2}$; see,
e.g. Galloway et al.\ 2008, Miller, Miller, \& Reynolds 2011).

We measure a disk blackbody temperature of $kT = 1.833(5)$~keV and a
normalization of $K = 29.0(3)$, a simple blackbody temperature and
normalization of $kT = 2.495(5)$~keV and $K = 0.0222(3)$, and a
power-law index and normalization of $\Gamma = 3.41(3)$ and $K =
1.98(4)~ {\rm photons}~ {\rm cm}^{-2}~ {\rm s}^{-1}$ at 1 keV.  For
``kerrdisk'', we fixed outer emissivity index to equal the inner
emissivity index, constrained the energy to the range possible for Fe
(6.40--6.97~keV), and measured the line centroid energy, spin
parameter, inclination, inner emission radius, and line flux.  We
measure a line centroid energy of $E = 6.97_{-0.01}$~keV (consistent
with H-like Fe XXVI), an emissivity index of $q = 3.5(1)$, a spin
parameter of $a = 0.17(2)$, an inner disk inclination of $i = 18(2)$
degrees, an inner radius of $r_{in} = 1.8(1)~r_{ISCO}$ ($10.6\pm 0.6~
r_{g}$), and a flux of $5.5\times 10^{-3}~ {\rm photons}~ {\rm
  cm}^{-2}~ {\rm s}^{-1}$.  This flux translates into a line
equivalent width of $W = 91(2)$~eV.

Fits with a relativistic line function are strongly required over a
Gaussian model for the emission line.  The best fits with the
``kerrdisk'' and ``diskline'' functions give $\chi^{2}/\nu =
1010.9/912$.  Instead modeling the line with a Gaussian corresponding
to any charge state of Fe (6.40--6.97~keV) gives $\chi^{2}/\nu =
1052.1/915$ (E = $6.53(2)$~keV, $\sigma = 0.44(2)$~keV).  Via an
F-test, a relativistic line function is therefore required at the
$5.3\sigma$ level of confidence.

It is also possible to test whether or not the line profile is
actually a composition of different plausible charge states.  Adding
two Gaussians from any charge state of Fe, constrained to have the
same width but allowed to have different strengths, gives a
significantly worse fit than a single relativistic line: $\chi^{2}/\nu
= 1266.1/915$.  This effectively eliminates the possibility that the
line profile arises in any form of diffuse gas far from the stellar
surface.

Using this phenomenological model, we measure an unabsorbed flux of
$6.1\times 10^{-9}~ {\rm erg}~ {\rm cm}^{-2} {\rm s}^{-1}$ in the
3.0--40.0~keV band, implying $1.5\times 10^{-8}~ {\rm erg}~ {\rm
  cm}^{-2} {\rm s}^{-1}$ in the 0.5--40.0~keV band.  The 3--40~keV
range is closer to that common in X-ray monitors and corresponds to
0.2--0.3~Crab.  The distance to Serpens X-1 is likely $7.7\pm 0.9$~kpc
(based on X-ray burst properties; see Galloway et al.\ 2008).  A
luminosity of $1.1(2) \times 10^{38}~ {\rm erg}~ {\rm s}^{-1}$ follows
from the inferred 0.5--40.0~keV flux.  It is likely that Serpens X-1
was accreting at a high fraction of its Eddington limit during our
observation.

Figure 1 shows clear hallmarks of disk reflection, even after fitting
the relativistic line, including an edge between 9--10~keV consistent
with H-like Fe XXVI, and a flux excess in the 10--20~keV range
consistent with a Compton back-scattering hump.  It is also clear in
Figure 1 that most of the flux capable of ionizing Fe comes from the
blackbody component.  To
self--consistently model the reflection, then, we employed a new
version of the well--known ``reflionx'' model (Ross \& Fabian 2005),
wherein a blackbody illuminated the disk instead of the power-law.

The ``kerrdisk'' line model was replaced by ``reflionx'',
convolved with the kernel of ``kerrdisk'', called ``kerrconv''.
Its parameters include inner and outer disk emissivity
indices ($q_{out}$ was fixed to $q_{in}$ in our fits), the inclination
of the inner disk, the spin parameter of the compact object, and the
inner and outer emitting radii (in units of the ISCO radius; $R_{out}
= 400$ was fixed in all cases).  The ``reflionx'' model variables
include the ionization parameter of the disk ($\xi = L/nr^{2}$), the
abundance of iron ($A_{Fe}$), the blackbody temperature (tied to that
of the independent blackbody), and the flux.

Table 1 lists the results of models for the time-averaged
spectrum of Serpens X-1, including relativistically--blurred disk
reflection (Models 1--18).  The spin of Serpens X-1 is unknown as kHz
QPOs and burst oscillations have not been detected.  For
completeness we considered fits with three spin values: $a = 0.0,
0.12, 0.24$.  The latter two values span the range of spin parameters
implied in other ``atoll'' sources (see, e.g. Galloway et al.\ 2008),
assuming $R_{NS} = 10$~km, $M_{NS} = 1.4~ M_{\odot}$, and $I_{NS} =
(2/5) M R^{2}$ as per a simple sphere.  Similarly, the iron abundance
of Serpens X-1 is unknown, so we explored fits with three values
(relative to solar): $A_{Fe} = 1.0, 2.0, 3.0$.  Last, in order to test
the sensitivity of the spectrum to the inner radius of the disk, we
considered fits with variable inner radii, and fits with
$r_{in} = 1.0~ r_{ISCO}$.

The best blurred reflection fits (Models 5, 11, 17) uniformly give
significantly better fits than models without reflection: an F-test
gives an $F = 96.3$, or a false-alarm probability of just $1.1\times
10^{-21}$.  Combined with the ability of the reflection models to
naturally model the 10--20~keV flux excess in Figure 1, this signals
that the full broad-band disk reflection spectrum has been detected in
Serpens X-1.

All of the models in Table 1 prefer very low inclinations; this
corresponds well with recent optical observations that independently
point an inclination of $10^{\circ}$ or less (Cornelisse et
al.\ 2013).  The results in Table 1 also clearly signal that the fits
are largely insensitive to the spin parameter of the neutron star.
This is consistent with expectations: even for $a = 0.24$, the ISCO
radius only changes marginally to $r_{ISCO} = 5.89 r_{g}$ (where
$r_{g} = GM/c^{2}$).

The data show a slight preference for an enhanced iron abundance in
Serpens X-1.  Given that the binary orbital period of Serpens X-1
appears to be just 2 hours (Cornelisse et al.\ 2013), it is possible
that the companion is evolved or degenerate, skewing e.g. Fe/O
abundance ratios.  Enhanced abundances are partly degenerate with the
normalization of the reflection component and its ionization.
Excellent fits are achieved regardless of the abundance, however, and
parameters of greater interest show no trend with abundance.

The models wherein the inner disk radius is allowed to float nominally
prefer a radius that is slightly larger than the ISCO; however, the
improvement in the fit statistic is not significant, and 90\% and/or
$3\sigma$ confidence errors include the ISCO radius.  Fits to the
time-averaged spectra may be characterized as being broadly consistent
with a disk extending to the ISCO.

We also examined the broad flux trends within the observation,
and identified a period of high flux for consideration (see Figure 3).
Total FPMA and FPMB spectra, background spectra, and responses were
created for ``GTI2'' in exactly the same manner as for the
time-averaged spectrum.  The results of fits to this segment of the
observation are also given in Table 1 (see Models 19--24). 

Figure 4 shows plots of $\Delta \chi^{2}$ versus $r_{in}$ for models 5
and 23, in order to illustrate the sensitivity of the spectra to the
inner extent of the disk.  Fits to the time-averaged (Model 5) and
high-flux (Model 23) spectra point to a disk that hovers close to the
ISCO.  Model 5 prefers $r_{in} > r_{ISCO}$, but Model 23
suggests that $r_{in} = r_{ISCO}$ is fully compatible with the data.
The disk continuum is less incisive than the reflection spectrum, but
it can serve as a broad consistency check.  The disk continua in Table
1 imply inner radii of 13--15~km (1.03--1.20~$r_{ISCO}$ or
6.2--7.2~$r_{g}$ for $M_{NS} = 1.4~ M_{\odot}$ and $a=0$) for a
canonical spectral hardening factor of 1.7 (Shimura \& Takahara 1995).

For $a=0$, the gravitational redshift implied by the fits is given by
$z + 1 = 1 / \sqrt{1 - 2GM/rc^{2}}$.  The best models for the
time-averaged spectra in Table 1 (e.g. Model 5) give $r = 1.3~
r_{ISCO} = 7.8~r_{g}$.  The neutron star must be smaller than the
inner disk radius, giving a limit of $z_{NS} \geq 0.16$.  Within the
high flux interval, even smaller inner disk radii are preferred, and
errors include $r = r_{ISCO}$.  If the disk extends to the ISCO,
$z_{NS} \geq 0.22$ and $R_ {NS} \leq 12.6$~km assuming $M_{NS} = 1.4~
M_{\odot}$.  The masses of neutron stars in low-mass X-ray binaries
are not known precisely, but $1.4~ M_{\odot}$ is a reasonable average
(Lattimer 2011).  If the neutron star surface itself truncates the
disk (and if any boundary layer is small in radial extent), or if the
effective reflection radius extends within the ISCO, then these limits
could be regarded as measurements.

\section{Discussion}
We have analyzed an initial observation of the neutron star low-mass
X-ray binary Serpens X-1.  Owing to the abilities of {\it NuSTAR}, an
Fe K line in the spectrum is revealed to be skewed at a high level of
statistical confidence, and the Compton back-scattering ``hump''
expected in disk reflection scenarios is detected in a neutron star
X-ray binary for the first time.  Both findings are robust since the
FPMA and FPMB detectors do not suffer from flux disortions due to
effects such as photon pile-up.  Clearly, broad Fe K lines in at least
some neutron star X-ray binaries arise in the inner disk, and can be
used to probe the stellar properties and inner accretion flow.

In contrast to black holes, an ISCO is not necessarily a feature of
accretion onto neutron stars.  The stellar radius could exceed its own
ISCO, and/or the stellar magnetic field can truncate the disk at much
larger radii.  The latter is certainly the case in typical X-ray
pulsars, but also in transient sources such as SAX J1808.4$-$3658
(Cackett et al.\ 2009) and IGR J17480$-$2446 in Terzan 5 (Miller et
al.\ 2011).  Fits to the robust data obtained with {\it NuSTAR}
indicate that the disk likely extends to the ISCO, or hovers just
above such a radius.  The high luminosity implied during this
observation makes it more likely that the disk is not impeded by a
magnetic field; this is also indicated by fits to a high--flux
interval of the observation.

If the neutron star radius in Serpens X-1 is small, this observation
represents the strongest observational indication of an ISCO in a
neutron star low-mass X-ray binary.  In this case, our fits require
$z_{NS} \geq 0.16$ (best fit), or $z \geq 0.22$ and $R_{NS} \leq
12.6$~km if $r = r_{ISCO}$, assuming $M_{NS} = 1.4~ M_{\odot}$.
If the star in Serpens X-1 is larger than its ISCO, the disk may then
be truncated by a the stellar surface and/or a thin boundary layer.
It is also possible that material in the inner disk may leak across
the ISCO and be reflective between the ISCO and stellar surface.  In
either of these two scenarios, the limits placed by our fits may be
characterized as measurements.  Some equations of state may
then be incompatible with our observations (see Lattimer 2011).

The statistically significant skewing of the line profile clearly
signals that dynamics and gravitation close to the neutron star drive
our results.  However, the results are not entirely
model--independent.  A disk reflection spectrum is required by the
data.  Although we achieved excellent fits with ``reflionx'',
different and/or future reflection and relativistic convolution models
may yield slightly different results.

\hspace{0.1in}

This work was supported under NASA Contract No. NNG08FD60C, and made
use of data from the NuSTAR mission, a project led by the California
Institute of Technology, managed by the Jet Propulsion Laboratory, and
funded by NASA.

\clearpage

\begin{figure}
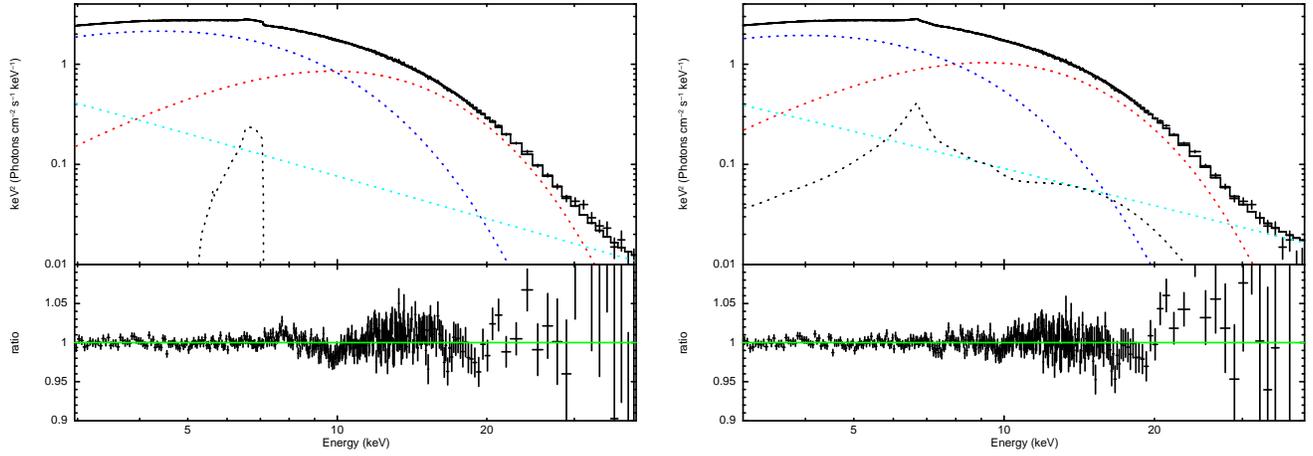

\includegraphics[scale=0.35,angle=-90]{fig1a.ps}
\includegraphics[scale=0.35,angle=-90]{fig1b.ps}
\figcaption[t]{\footnotesize The {\it NUSTAR} 3--40~keV spectrum of
  Serpens X-1 is shown above, with the FPMA and FPMB detectors
  combined.  {\it Left}: The best-fit phenomenological spectral model
  is shown here.  The model consists of disk blackbody (blue), simple
  blackbody (red), power-law (cyan), and relativistic disk line
  (black) components.  The edge feature between 9--10~keV is
  consistent with the K edge of Fe XXVI, and the flux excess in the
  10--20 keV band is consistent with the Compton back-scattering
  ``hump'' expected from disk reflection.  {\it Right}: The best
  relativistically--blurred disk reflection fit (Model 5 in Table 1)
  is shown here.  The line function has been replaced with a version
  of ``reflionx'' suited to blackbody illumination of the disk, and
  convolved with the ``kerrconv'' function.  A significantly better
  fit is achieved.}
\end{figure}
\medskip

\begin{figure}
\includegraphics[scale=0.7]{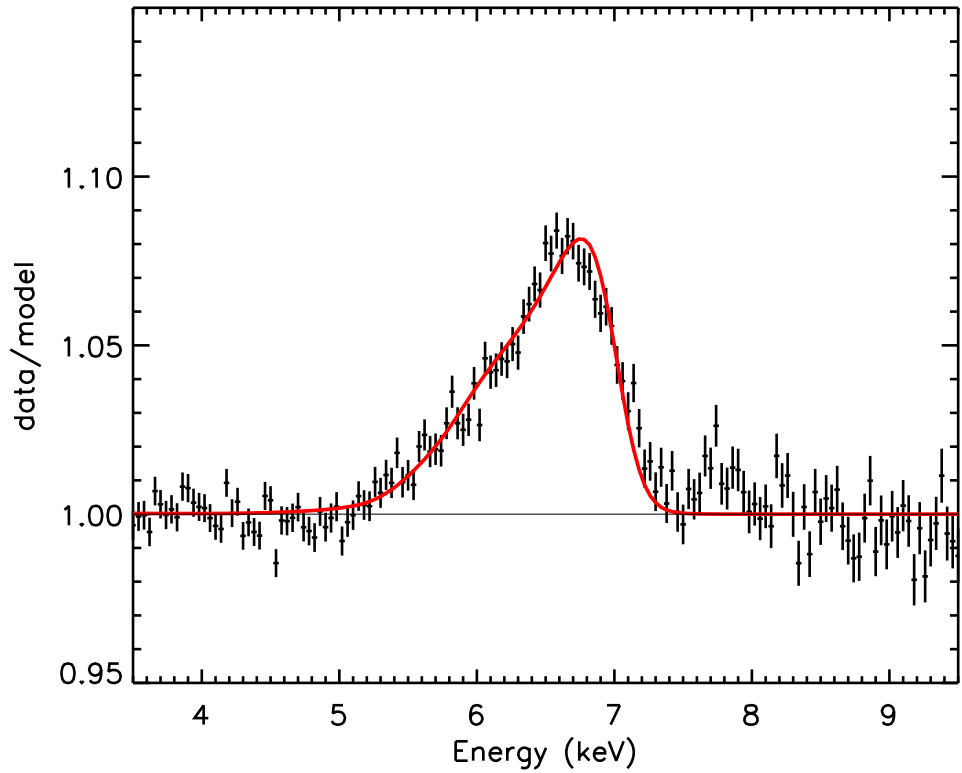}
\figcaption[t]{\footnotesize The relativistic Fe K disk line in the
  combined FPMA and FPMB spectrum of Serpens X-1 is shown above.  The best
  three--component continuum model was fit over the 3--40~keV band,
  ignoring the 4.5--7.5 keV range to avoid biasing the continuum fit.
  The best ``kerrdisk'' model, derived from fitting the full bandpass
  including the Fe K region, is shown in red.  The spectrum was
  binned for visual clarity only.}
\end{figure}
\medskip

\begin{table}[t]
\caption{Relativistically-blurred Disk Reflection Models}
\begin{scriptsize}
\begin{center}
\begin{tabular}{lllllllllllllllll}
\tableline
Spectrum & Model & $kT_{disk}$ & $K_{disk}$ & $kT_{bb}$ & $K_{bb}$ & $\Gamma$ & $K_{pow}$ & $a$ & $q$ & $r_{in}$ & $i$ & $\xi$ & $A_{Fe}$ & $K_{refl}$ & $\chi^{2}/\nu$ \\
   ~     &  ~    & (keV)     &  ~    &   (keV)   &  ($10^{-2}$) &    ~     &   ~      &  ~  &  ~  & (ISCO) & (deg.) & (erg cm ${\rm s}^{-1}$) & ~ & ($10^{-11}$)  & ~  \\
\tableline
time-avg & 1 & 1.60(1) & 45(2) &   2.33(1) & 2.52(8) & 3.17(3) & 1.3(1) &              0.0* & 2.21(4) & 1.2(1) & $8_{-8}^{+2}$ & 370(40)   & 1.0*  & 4.5(1) & 995.1/914 \\
time-avg & 2 & 1.61(3) & 44(3) & 2.33(2) & 2.5(1) & 3.18(5) & 1.4(2) &                  0.0* & 2.19(3) & 1.0*   & $8_{-8}^{+2}$ & 330(80)   & 1.0*  & 5(2) & 998.6/915 \\
time-avg. & 3 & 1.61(2)  &  41(2)  &  2.33(1)  &  2.63(6)  &   3.21(3)   &  1.5(1)  &  0.0*  & 2.26(4)  & 1.3(1)  & $8_{-8}^{+2}$ & 220(60) & 2.0* & 5(2) & 962.3/913 \\
time-avg. & 4 & 1.64(3) & 41(3)     & 2.33(2)   & 2.6(1)    & 3.22(4)     & 1.5(1)     & 0.0* & 2.23(3) & 1.0* & $8_{-8}^{+2}$ & 300(90)   & 2.0* &  6(3) & 966.0/915 \\
time-avg. & 5  & 1.64(2) & 41(3) & 2.33(2) & 2.65(5) & 3.23(4) & 1.5(1) &              0.0* & 2.29(3) & 1.3(1) & $8_{-8}^{+2}$ & 200(20)   & 3.0* & 7(2)  & 950.9/914 \\
time-avg & 6 & 1.64(2) & 41(3) &    2.33(2) & 2.70(5) & 3.24(4) & 1.6(1) &              0.0* & 2.27(3) & 1.0*   & $8_{-8}^{+2}$ & 270(90)   & 3.0*  & 6(2) & 955.9/915 \\
\tableline
time-avg & 7 & 1.63(3) & 42(2) &  2.34(1)  & 2.37(8) & 3.21(2) & 1.48(3) &              0.12* &  2.24(4)  & 1.25(3) & $8_{-8}^{+2}$ & 240(30) & 1.0* & 6.0(9) & 998.4/914 \\
time-avg & 8 & 1.61(3) &  43(3)  &  2.33(2)  & 2.5(1) & 3.21(3)  & 1.5(1)  &              0.12* &  2.20(4) & 1.0*    & $8_{-8}^{+2}$ & 340(90) & 1.0* & 5(2)   & 998.7/915 \\ 
time-avg & 9 & 1.63(2) & 41(2) & 2.33(1) & 2.6(1)  &  3.21(3) & 1.5(1)  &              0.12* &   2.24(4)  & 1.3(2)  & $8_{-8}^{+2}$ & 250(60) & 2.0* & 7(3)   & 962.4/914 \\
time-avg & 10 & 1.64(3)  & 41(3) & 2.33(2) & 2.61(9) &  3.23(4) &  1.6(2)  &              0.12* &  2.22(3)  & 1.0*    & $8_{-8}^{+2}$ & 300(90) & 2.0* & 6(2)   & 966.6/915 \\
time-avg & 11 & 1.64(2) & 40(1) & 2.33(1)  &  2.69(3) & 3.22(2) &  1.51(5) &              0.12* &  2.27(3) & 1.35(3) & $8_{-8}^{+2}$ & 210(20) & 3.0* & 6.1(6) & 950.9/914 \\
time-avg & 12 & 1.64(3) & 40(3) & 2.33(2)  & 2.69(9) & 3.24(4) & 1.6(2)   &              0.12* &   2.26(3)  & 1.0*    & $8_{-8}^{+8}$ & 280(90) & 3.0* & 5(3)   & 957.1/915 \\
\tableline
time-avg & 13 & 1.64(1) & 42(2) & 2.35(1) & 2.5(2) & 3.20(2) & 1.5(1) &               0.14* & 2.22(3) & 1.4(1) & $8_{-8}^{+2}$ & 290(40) & 1.0* & 5.2(4) & 995.8/914 \\
time-avg & 14 & 1.61(3) & 44(3) & 2.34(2)  & 2.5(2) & 3.19(4) & 1.5(2) &               0.14* & 2.19(3) & 1.0*   &$8_{-8}^{+4}$ &  330(90) & 1.0* & 5(2) & 1000.0/915 \\
time-avg & 15 & 1.63(1) & 41(2) & 2.33(1) & 2.6(1) & 3.22(2) & 1.47(6) &              0.14* & 2.25(3) & 1.4(1) & $8_{-8}^{+2}$ & 220(60) & 2.0* & 7(2) & 962.2/914 \\
time-avg & 16 & 1.64(3) & 41(3) &  2.33(2) & 2.7(2) & 3.24(4) & 1.6(2) &               0.14* & 2.23(3) & 1.0*   & $8_{-8}^{+2}$ & 300(90) & 2.0* & 6(3) & 967.6/915 \\
time-avg & 17 & 1.65(1) & 41(1) & 2.31(3) & 2.69(4) & 3.20(2) & 1.51(4) &             0.14* & 2.27(3) & 1.41(4) & $8_{-8}^{+2}$ & 210(30) & 3.0* & 6(1) & 951.0/914 \\
time-avg & 18 & 1.65(2) & 41(3) & 2.33(1)  & 2.7(1) & 3.25(4) & 1.6(2) &              0.14* & 2.27(3) &  1.0*   & $8_{-8}^{+8}$ & 270(90) & 3.0* & 5(2) & 958.1/915 \\ 
\tableline
high-flux & 19 & 1.75(5) & 37(3) & 2.40(2) & 2.6(2) & 3.30(5) & 1.7(3) &              0.00* & 2.21(7) &  $1.2_{-0.2}^{+0.4}$ & $8_{-8}^{+3}$  & 380(90)  & 1.0* & 5(2)   & 936.7/914  \\ 
high-flux & 20  & 1.73(5) & 36(4) & 2.49(4) & 2.6(2) & 3.31(7) & 1.8(3) &              0.00* & 2.21(6) &  1.0* &             $8_{-8}^{+4}$  & 360(80)  & 1.0* & 5(2)   & 937.1/915  \\
high-flux & 21 & 1.73(6) & 36(3) & 2.39(3) & 2.6(3) & 3.35(9) & 1.8(3) &              0.00* & 2.26(6) &  $1.1_{-0.1}^{+0.4}$ & $8_{-8}^{+3}$  & 310(90)  & 2.0* & 6(3)   & 928.8/914  \\
high-flux & 22  & 1.75(6) & 35(5) & 2.38(3) & 2.6(3) & 3.33(7) & 1.8(3) &              0.00* & 2.25(6) &  1.0* &             $8_{-8}^{+4}$  & 330(90)  & 2.0* & 6(3)   & 929.6/915  \\
high-flux & 23 & 1.73(4) & 35(4) & 2.38(2) & 2.7(2) & 3.33(7) & 1.8(3) &              0.00* & 2.31(7) &  $1.2_{-0.2}^{+0.3}$ &  $8_{-8}^{+4}$ & 330(90)  & 3.0* & 5(2)   & 926.4/914  \\
high-flux & 24  & 1.76(6) & 34(4) & 2.39(4) & 2.7(3) & 3.35(8) & 1.9(4) &              0.00* & 2.29(5) &  1.0* &              $8_{-8}^{+5}$ & 340(90)  & 3.0* & 6(3    & 927.5/915  \\  

\tableline
\end{tabular}
\vspace*{\baselineskip}~\\ \end{center} 
\tablecomments{The results of fits with relativistically-blurred disk
  reflection components (``reflionx'' convolved with ``kerrconv'') are
  presented here.  The disk emissivity is measured by the parameter
  $q$ ($J ~ r^{-q}$).  Parameters fixed at a given value are marked
  with an asterisk.  The ionization of the disk is measured by the
  ionization parameter $\xi$.  All fits were made assuming an
  equivalent neutral hydrogen column density of $N_{H} = 4.0\times
  10^{21} {\rm cm}^{-2}$ as per Dickey \& Lockman (1990), via the
  ``tbabs'' model.  All errors quoted in the table are $1\sigma$
  uncertainties.}
\vspace{-1.0\baselineskip}
\end{scriptsize}
\end{table}
\medskip

\clearpage

 \begin{figure}
 \vspace{3.0in}
 \includegraphics[scale=1.0]{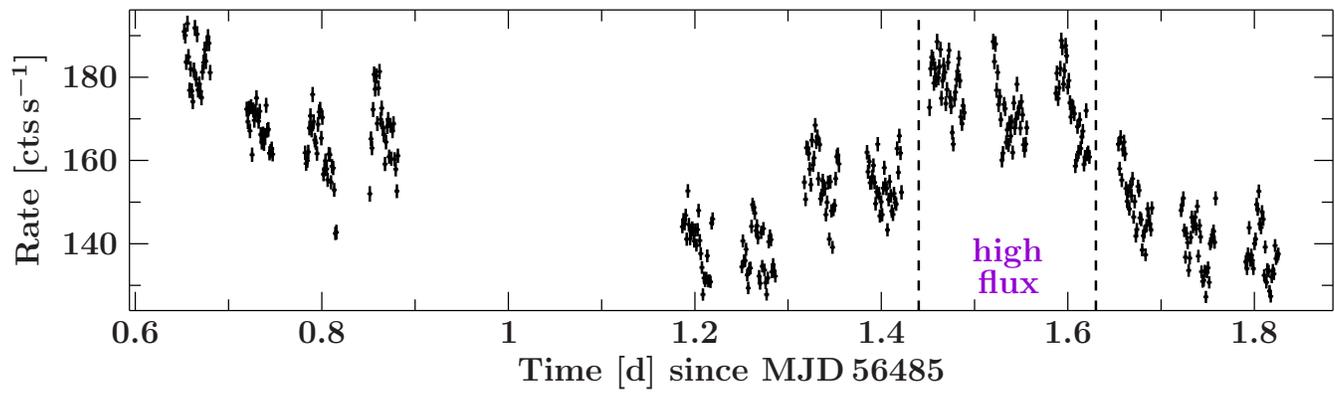}
\figcaption[t]{\footnotesize The {\it NuSTAR}
   light curve of Serpens X-1.  The interval defined as ``high flux'' was
  selected for independent spectral analysis for comparison (see Table
  1).}
 \end{figure}
 \medskip


 \begin{figure}
 \includegraphics[scale=0.6]{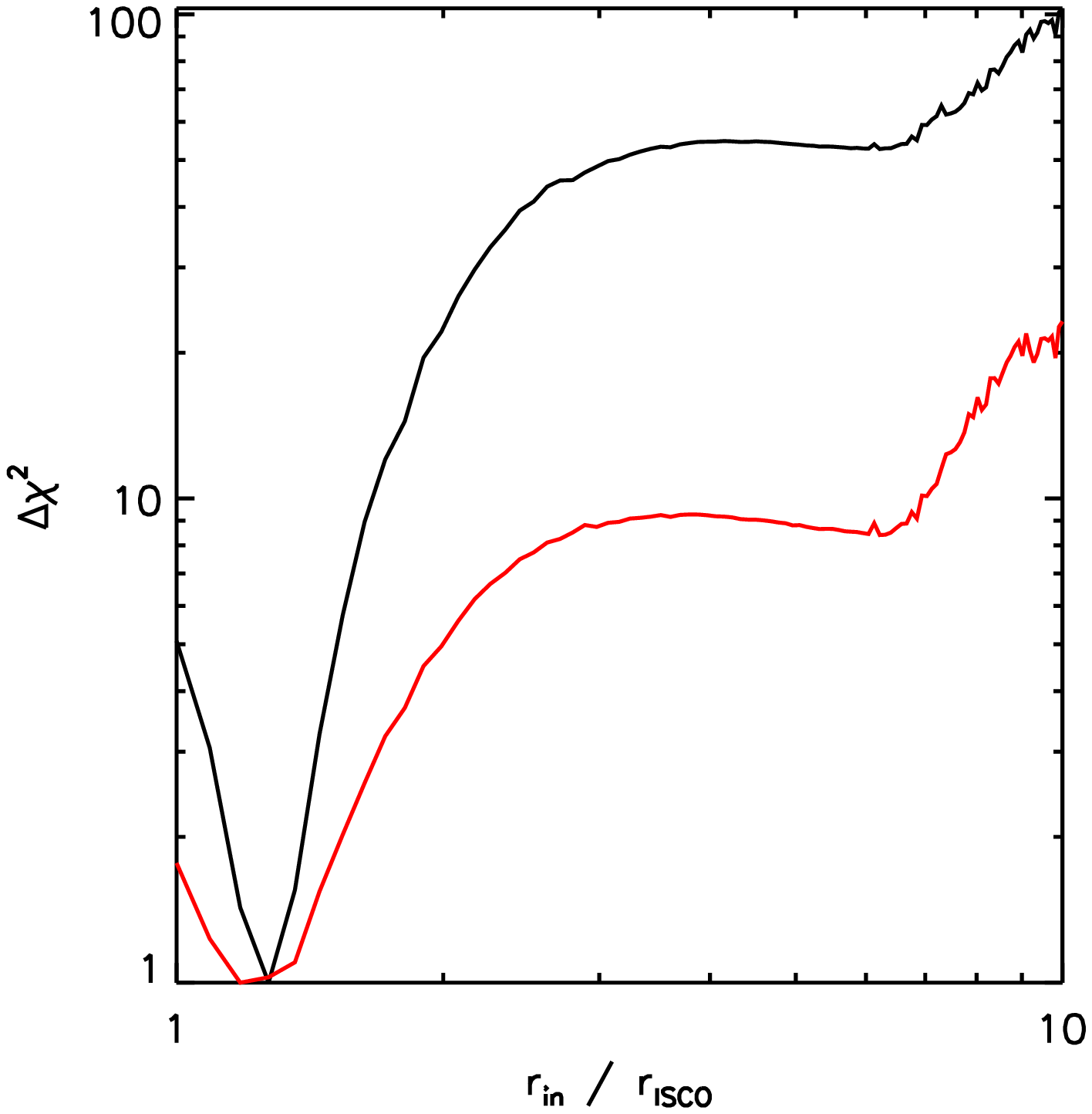}
 \figcaption[t]{\footnotesize The figure above displays the change in
  the fitting statistic, $\Delta \chi^{2}$, versus the inner disk
  radius measured via two blurred reflection fits.  Model 5, which is
  a fit to the time-averages spectrum of Serpens X-1, is shown in
  black.  Model 23, which is a fit to the high-flux phase of the
  observation (see Section 3), is shown in red.  Both models, like
  others detailed in Table 1, strongly prefer a disk that is close to
  a likely ISCO, and are statistically consistent with the disk
  extending to the ISCO itself.  For plotting clarity, a constant of
  1.0 was added to all $\Delta \chi^{2}$ values.}
 \end{figure}
 \medskip


\begin{references}

\reference{} Arnaud, K. A., and Dorman, B., 2000, XSPEC is available
via the HEASARC on-line service, provided by NASA/GSFC

\reference{} Bhattacharyya, S., \& Strohmayer, T., 2007, ApJ, 64, L103

\reference{} Boutloukos, S., Miller, M., Lamb, K., 2010, ApJ, 720, L15

\reference{} Brenneman, L., \& Reynolds, C. S., 2006, ApJ, 652, 1028

\reference{} Cackett, E. M., Miller, J. M., Bhattacharyya, S.,
Grindlay, J. E., Homan, J., van der Klis, M., Miller, M. C.,
Strohmayer, T. E., Wijnands, R., 2008, ApJ, 674, 415

\reference{} Cackett, E. M., Miller, J. M., Ballantyne, D. R., Barret,
D., Bhattacharyya, S., Boutelier, M., Miller, M. C., Strohmayer,
T. E., Wijnands, R., 2010, ApJ, 720, 205

\reference{} Cackett, E. M., Miller, J. M., Reis, R., C., Fabian,
A. C., Barret, D., 2012, ApJ, 755, 27

\reference{} Churazov, E., Gilfanov, M., Forman, W., Jones, C., 1996, ApJ, 471, 673

\reference{} Cornelisse, R., Casares, J., Charles, P. A., \& Steeghs, D., 2013, MNRAS, 432, 1361

\reference{} D'Ai, A., Iaria, R., Di Salvo, T., Matt, G., Robba, N. R., 2009, ApJ, 693, L1

\reference{} Demorest, P. B., Pennucci, T., Ransom, S. M., Roberts,
M. S., E., Hessels, J. W. T., 2010, Naure, 467, 1081

\reference{} Diaz Trigo, M., Sidoli, L., Boirin, L., Parmar, A. N., 2012, A\&A, 543, 50

\reference{} Dickey, J. M., \& Lockman, F., J., 1990, ARA\&A, 28, 215 

\reference{} Di Salov, T., D'Ai, A., Iaria, R., Burderi, L., Dovciak, M., KAras, V., Matt, G., Papitto, A., Piraino, S., Riggio, A., Robba, N. R., Santangelo, A., 2009, MNRAS, 398, 2022

\reference{} Egron, E., Di Salvo, T., Motta, S., Burderi, L., Papitto, A., Duro, R., D'Ai, A., Riggio, A., Belloni, T., Iaria, R., Robba, N., Paraino, S., Santangelo, A., 2013, A\&A, 550, 5

\reference{} Galloway, D. K., Muno, M. P., Hartman, J. M., Psaltis, D., Chakrabarty, D., 2008, ApJS, 179, 360

\reference{} Guillot, S., Servaillat, M., Webb, N., Rutledge, R. E., 2013, ApJ, 772, 7

\reference{} Guver, T., Ozel, F., Cabrera-Lavers, A., Wroblewski, P., 2010, ApJ, 712, 964

\reference{} Guver, T., Psaltis, D., Ozel, F., 2012a, ApJ, 747, 76

\reference{} Guver, T., Ozel, F., Psaltis, D., 2012b, ApJ, 747, 77

\reference{} Harrison, F. A., et al., 2013, ApJ, 770, 103

\reference{} Iaria, R., D'Ai, A., Di Salvo, T., Robba, N. R., Riggio, A., Papitto, A., Burderi, L., 2009, A\&A, 505, 1143

\reference{} Lattimer, J. M., 2011, Ap\&SS, 336, 67

\reference{} Lattimer, J. M., \& Prakash, M., 2007, PhR, 442, 109

\reference{} Lin, D., Homan, J., \& Remillard, R. A., 2010, ApJ, 719, 1350

\reference{} London, R. A., Taam, R. E., \& Howard, W. M., 1986, ApJ, 306, 170

\reference{} Magdziarz, P., \& Zdziarski, A. A., 1995, MNRAS, 273, 837

\reference{} Miller, J. M., 2007, ARA\&A, 45, 441

\reference{} Miller, J. M., Reynolds, C. S., Fabian, A. C., Miniutti,
G., Gallo, L. C., 2009, ApJ, 697, 900

\reference{} Miller, J. M., D'Ai, A., Bautz, M. W., Bhattacharyya, S.,
Burrows, D. N., Cackett, E. M., Fabian, A. C., Freyberg, M. J.,
Haberl, F., Kennea, J., Nowak, M. A., Reis, R. C., Strohmayer, T. E.,
Tsujimoto, M., 2010, ApJ, 724, 1441

\reference{} Miller, J. M., Miller, M. C., \& Reynolds, C. S., 2011, ApJ, 731, 5

\reference{} Miller, J. M., Parker, M. L., Fuerst, F., Bachetti, M., Harrison, F. A., Barret, D., Boggs, S. E., Chakrabarty, D., Chistensen, F. E., Craig, W. W., Fabian, A. C., Grefenstette, B. W., Hailey, C. J., King, A. L., Stern, D. K., Tomsick, J. A., Walton, D. J., Zhang, W. W., 2013, ApJ, in press, arxiv:1308.4669 

\reference{} Mitsuda, K., Inoue, H., Koyama, K., Makishima, K., Matsuoka, M., Ogawara, Y., Suzuki, K., Tanaka, Y., Shibazaki, N., Hirano, T., 1984, PASJ, 36, 741

\reference{} Ng, C., Diaz Trigo, M., Cadolle Bel, M., \& Migliari, S.,
2010, A\&A, 522, 96

\reference{} Papitto, A,. Di Salvo, T., D'Ai, A., Iaria, R., Burderi, L., Riggio, A., Manna, M. T., \& Robba, N. R., 2009, A\&A, 493, L39

\reference{} Risaliti, G., Harrison, F. A., Madsen, K. K., Walton, D. J., Boggs, S. E., Christensen, F. E., Craig, W. W., Grefenstette, B. W., Hailey, C. J., Nardini, E., Stern, D., Zhang, W. W., 2013, Nature, 494, 449

\reference{} Reynolds, C. S., \& Fabian, A. C., 2008, ApJ, 675, 1048

\reference{} Ross, R. R., \& Fabian, A. C., 2005, MNRAS, 358, 211

\reference{} Shafee, R., McKinney, J. C., Narayan, R., Tchekhovskoy,
A., Gammie, C. F., McClintock, J. E., 2008, ApJ, 687, L25

\reference{} Shimura, T., \& Takahara, F., 1995, ApJ, 445, 780

\reference{} Steiner, A., Lattimer, J., Brown, E., 2013, ApJ, 765, L5

\reference{} Tomsick, J., et al., 2013, ApJ, submitted.

\reference{} van der Klis, M., 2006, in ``Compact Stellar X-ray
Sources'', ed. W. H. G. Lewin \& M. van der Klis (Cambridge University
Press), 39

\reference{} Verner, D. A., Ferland, G. J., Korista, K. T., \&
Yakovlev, D. G., 1996, ApJ, 465, 487

\reference{} Wilms, J., Allen, A.,  McCray, R., 2000, ApJ, 542, 914

\end{references}
\end{document}